# Sequence Evolution, Structure and Dynamics of Transmembrane Proteins: Rhodopsin

J. C. Phillips

Dept. of Physics and Astronomy

Rutgers University, Piscataway, N. J., 08854

Abstract

Rhodopsin is a G-protein coupled receptor found in retinal rod cells, where it mediates monocrhromatic vision in dim light. It is one of the most studied proteins with thousands of reviewed entries in Uniprot. It has seven transmembrane segments, here examined for their hydrophobic character, and how that has evolved from chickens to humans. Elastic features associated with Proline are also discussed. Finally, differences between rhodopsin and cone opsins are also discussed.

1. Introduction

Improvement of protein function by evolution (natural selection) is expected on general grounds, but even with the modern database positive proof has remained a difficult problem for theory. While the static structural data bank is widely used, applications to evolution nearly always encounter problems. These are inevitable, not only because the static three-dimensional structures involving thousands of atoms are complex, but also because "evolution operates pragmatically: structures are not the target of selection, functions are" [1]. In a number of earlier papers we showed that evolution of non-transmembrane proteins can be described accurately by sequence analysis using phase-transition theory [2,3]; a recent example of great general interest is the evolution of COVID spikes and COVID contagiousness [4,5].

Non-transmembrane proteins are simpler because they function primarily through small changes in shape produced by small relative shifts of domains. A good first approximation treats these shifts in terms of a common length scale W, which can be found from overall evolutionary changes. This method could fail for transmembrane proteins, where the TM segments and outside loops are unevenly spaced. In practice structural experiments do not resolve structure, because the TM proteins can be studied only in the absence of membranes. Here we will combine studies of TM protein sequences with some general ideas about the structural differences between TM segments and outside loops to obtain new insights into TM protein functions.

It has been customary to describe the sequence evolution of TM proteins using BLAST online, which corresponds to W = 1. Here we examine a model transmembrane protein, rhodopsin. The rhodopsin structure features a seven-transmembrane (7TM) helix core architecture with three loop regions on both the extracellular and the cytoplasmic side of the membrane, with ~350 amino acids. We find that evolutionary features can still be distinguished using hydropathic shape changes with sliding window sizes W ≥ 3, but that these features are much weaker for W = 1. Moreover, because so much is already known about rhodopsin dynamics, we are able to discuss TM elastic interactions and show how these are supported hydrodynamically.

2. Methods

The methods used in this paper are drawn from a wide range of disciplines. These have been discussed in our earlier articles [4. 5]. A summary of them is available from arXiv [6]. The summary involves a minimum of technical background. Our earlier analysis relied on Ψ(aa,W) hydropathic profiles, where Ψ(aa) measures the hydropathicity of each amino acid. Small changes in protein shapes are often driven by waves in water films. These water waves have been averaged linearly over sliding windows of width W. (Data processing using sliding window algorithms is a general smoothing and sorting technique discussed online.) Often the best values of W are large (~25) for non-transmembrane proteins, reflecting lengths of dynamical domains.

3. Results

Our previous studies involved two thermodynamically different hydropathic scales: the KD scale [7], for first-order transitions, and the MZ scale [8], for second-order transitions. Most evolutionary changes in protein shape are small and spread over domains, so the MZ scale usually described evolution best. In advance, one might have expected each scale to yield interesting results, because the distinction between outside loops and inside TM segments is first order thermodynamically, while each is lengthy. Overall changes in hydropathic protein shape are measured most simply by variances of Ψ(aa,W) hydropathic profiles. These are shown in Fig. 1 for the chicken/human variance ratios for the MZ and KD scales, together with the BLAST result.

Profiles of Ψ(aa,W) for the human and chicken sequences for the MZ scale are shown in Fig. 2. Fig. 3 shows the human MZ profiles for three values of W. Fig. 4 compares MZ and KD human profiles for W = 9, where MZ and KD scales may be equally informative [4,5].

Discussion

Structural knowledge of rhodopsin is summarized on Uniprot, for instance human P08100. Structural studies, including molecular dynamics simulations, have been reviewed [9]. More recently, [10] discussed the effects of Prolines "in guiding large-scale conformational changes The hallmark of the active state of rhodopsinis the outward rotation of the intracellular end of TM helix H6. Coupled with H6 motion are changes in the orientations of the adjacent helices H5 and H7. Helices H5, H6 and H7 each contain a proline residue in the middle of the TM sequence: Pro215$^{5.50}$, Pro267$^{6.50}$ and Pro303$^{7.50}$, respectively". The centers of the hydrophobic regions listed in Table 1 (216, 259, and 305) agree well with the centered Pro sites. The agreement is better for W = 9 than for W = 5 or 15. Also, there is a double Proline in TM He4 {170,171}. This doublet can lock H4, and effectively isolate H5, H6, and H7 from thermal noise from the N half of rhodopsin [11].

Rhodopsins are placed in retinal rods, while Cone opsins are G-protein coupled receptors of cone outer segments in the vertebrate retina. While similar in sensitivity to rhodopsin, MW-opsin ran down less in response to repeated flashes and had faster kinetics [12]. The molecular basis for this improved performance is displayed in Fig. 5. The profiles are similar, but with one large difference, in H7. Rhodopsin contains three small hydrophobic peaks, which could compete for H&, whereas MW-opsin has only one.

Normally (Uniprot) TM segments are identified by combining helical segments with hydrophobic clusters. This works well here for most of the rhodopsin TM segments, especially 1-4. The profiles shown in the Figures reveal some new features. In these profiles TM segments 5 and 6 are usually the most hydrophobic, but Table 1 shows something unexpected. The entire region including TM segments 5 and 6 is helical (including the hydrophilic valley labelled HV in Fig. 4). Moreover, HV is the deepest of the hydrophilic loops connecting TM segments. This extended helical structure provides elastic stability and flexibility between H5 and H6, thus supporting their cooperative dynamics. Combined stability and flexibility are characteristic of critical points [2,3].

With regard to H7, there is no strong hydrophobic peak in this region; instead, there are three weak peaks. This is the region where retinal attachment occurs, and it is possible that retinal activation is facilitated by competing interactions of these three weak peaks.

We can also compare Pro sites. Overall BLAST shows 44% identities, while Pro is unevenly conserved; there are 11 conserved Pro sites, A: 7 in Rhodopsin and not in MW-opsin, and B: 4 in the converse. The 5 sites connected with TM segments, 215, 267, 303, 170,171, are conserved. Among the nonconserved A sites, there is a small cluster of 3 very near the N-terminal (center, 17), where Rhodopsin has beta strands. There is also a small cluster of 3 in B (center 43), which could be associated with beta strands in MW-opsin. Finally, the sequence of LW-opsin is also known. It has 96% identity with MW opsin, and all Pro's are conserved. LW is slightly more hydrophilic, enabling it to respond better in the longer wave red.

Availability of Data and Materials

The datasets analyzed during the current study are available in the Uniprot repository, [https://www.uniprot.org/].

# References

1. Greslehner, GP "What do molecular biologists mean when they say 'structure determines function'?" http://philsci-archive.pitt.edu (2018).
2. Munoz, MA Colloquium: Criticality and dynamical scaling in living systems. Rev. Mod. Phys. **90**, 031001 (2018).
3. Daniels, BC Kim, H Moore, D et al. Criticality distinguishes the ensemble of biological regulatory networks. Phys. Rev. Lett. **121**, 138102 (2018).
4. Phillips, JC Synchronized attachment and the Darwinian evolution of Coronaviruses CoV-1 and CoV-2. arXiv2008.12168 (2020); Phys. A **581**, 126202 (2021).

5. Phillips, JC From Omicron to Its BA.5 Subvariant: A New Twist in Coronavirus Contagiousness Evolution. arXiv 2208.04956 (2022).
6. Phillips, J.C. How Life Works: Darwinian evolution of proteins. arXiv 2105.12727 (2021).
7. Kyte, J Doolittle, RF A simple method for displaying the hydropathic character of a protein. **J. Mol. Biol. 157,** 105-132 (1982).
8. Moret, MA Zebende, GF Amino acid hydrophobicity and accessible surface area. Phys. Rev. E **75**, 011920 (2007).
9. Smith, SO Structure and activation of the visual protein rhodopsin. Annu. Rev. Biophys. **39**, 309-328 (2010).
10. Kimata, N., Pope, A., Sanchez-Reyes, O. *et al.* Free backbone carbonyls mediate rhodopsin activation. Nat Struct Mol Biol **23**, 738–743 (2016).
11. Pallesen, J Wang, N Corbett, KS et al. Immunogenicity and structures of a rationally designed prefusion MERS-CoV spike antigen. Proc. Natl. Acad. Sci. U.S.A. **114**, E7348–E7357 (2017).
12. Berry, M.H., Holt, A., Salari, A. *et al.* Restoration of high-sensitivity and adapting vision with a cone opsin. Nat Commun **10**, 1221 (2019).

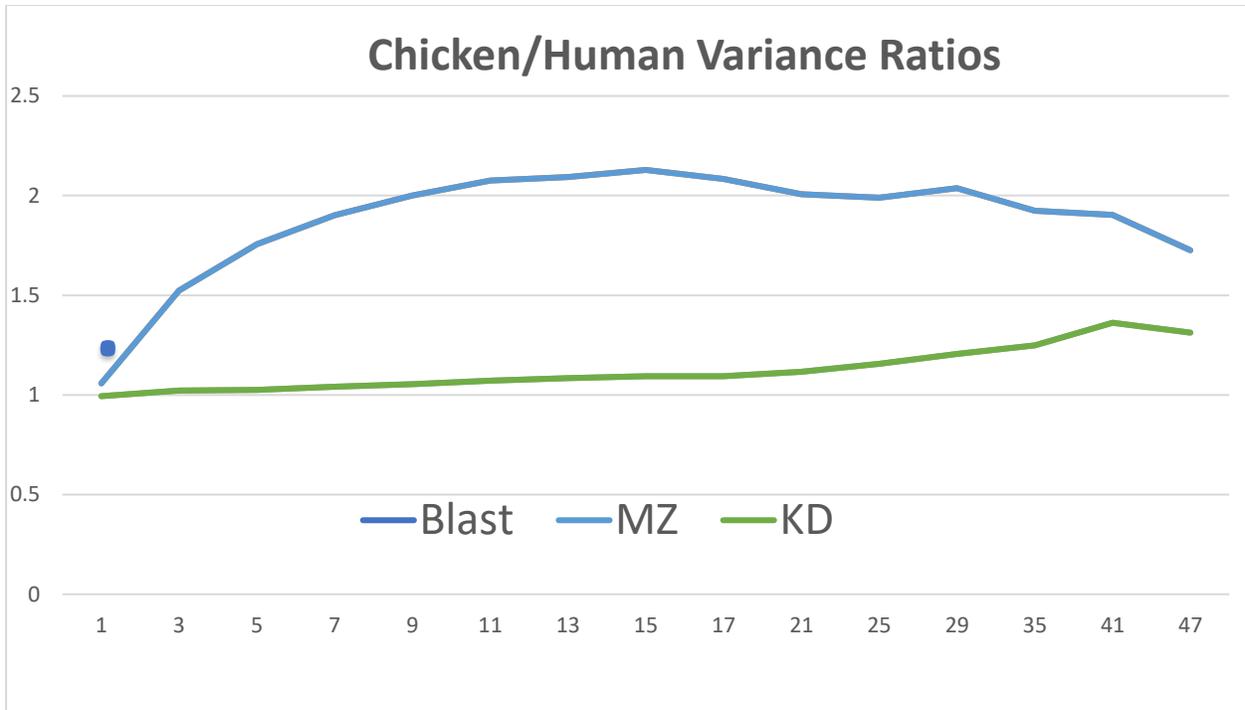

**Fig.1.** The variance ratios of Ψ(aa,W) for a range of W values from W = 1 to W = 47. Also shown is the BLAST ratio, which corresponds to W = 1. The effects of evolution are largest with the MZ scale over a wide range of W.

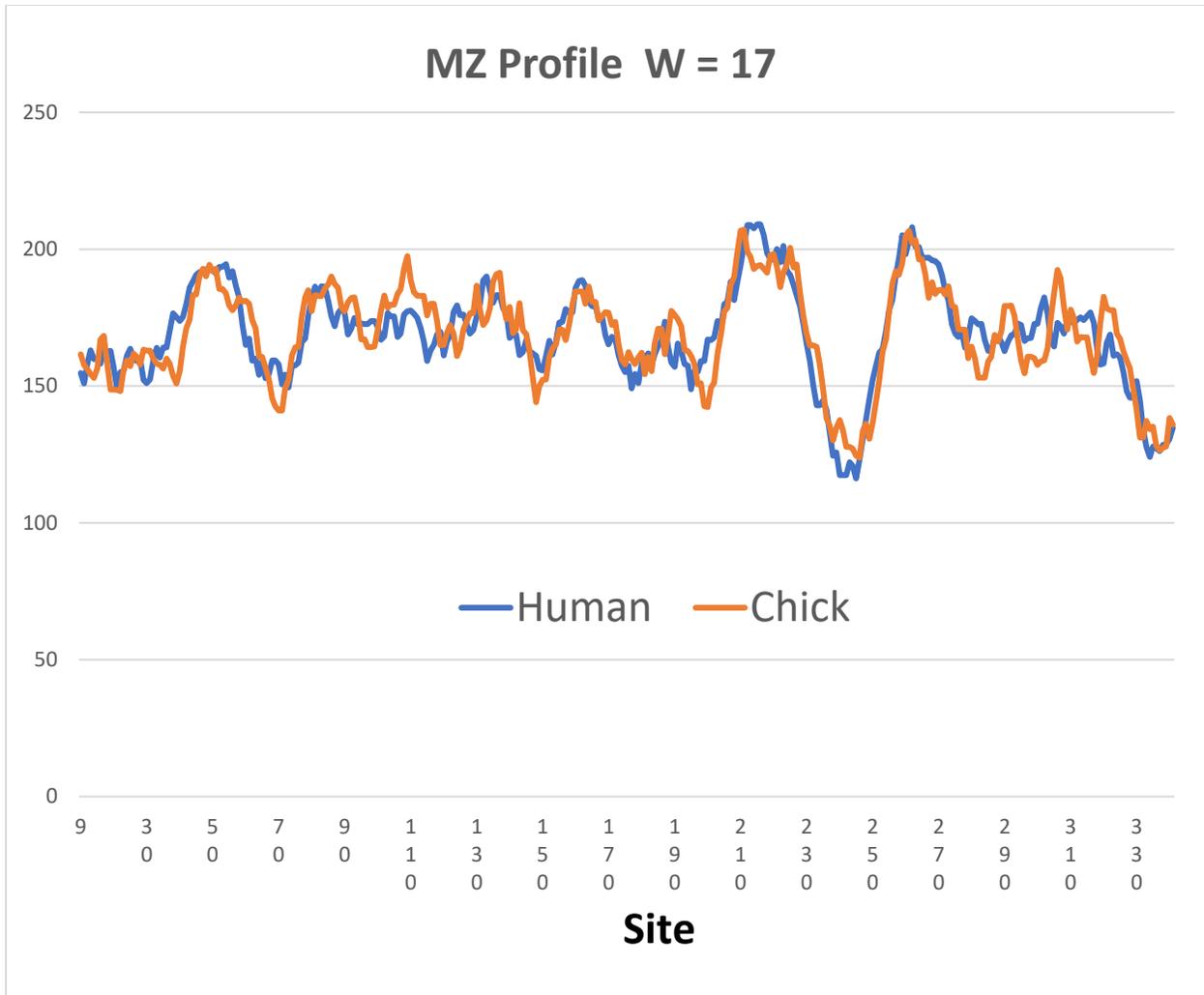

Fig. 2. Overall the evolutionary differences between chicken and human rhodopsin Ψ(aa,W) profiles are small.

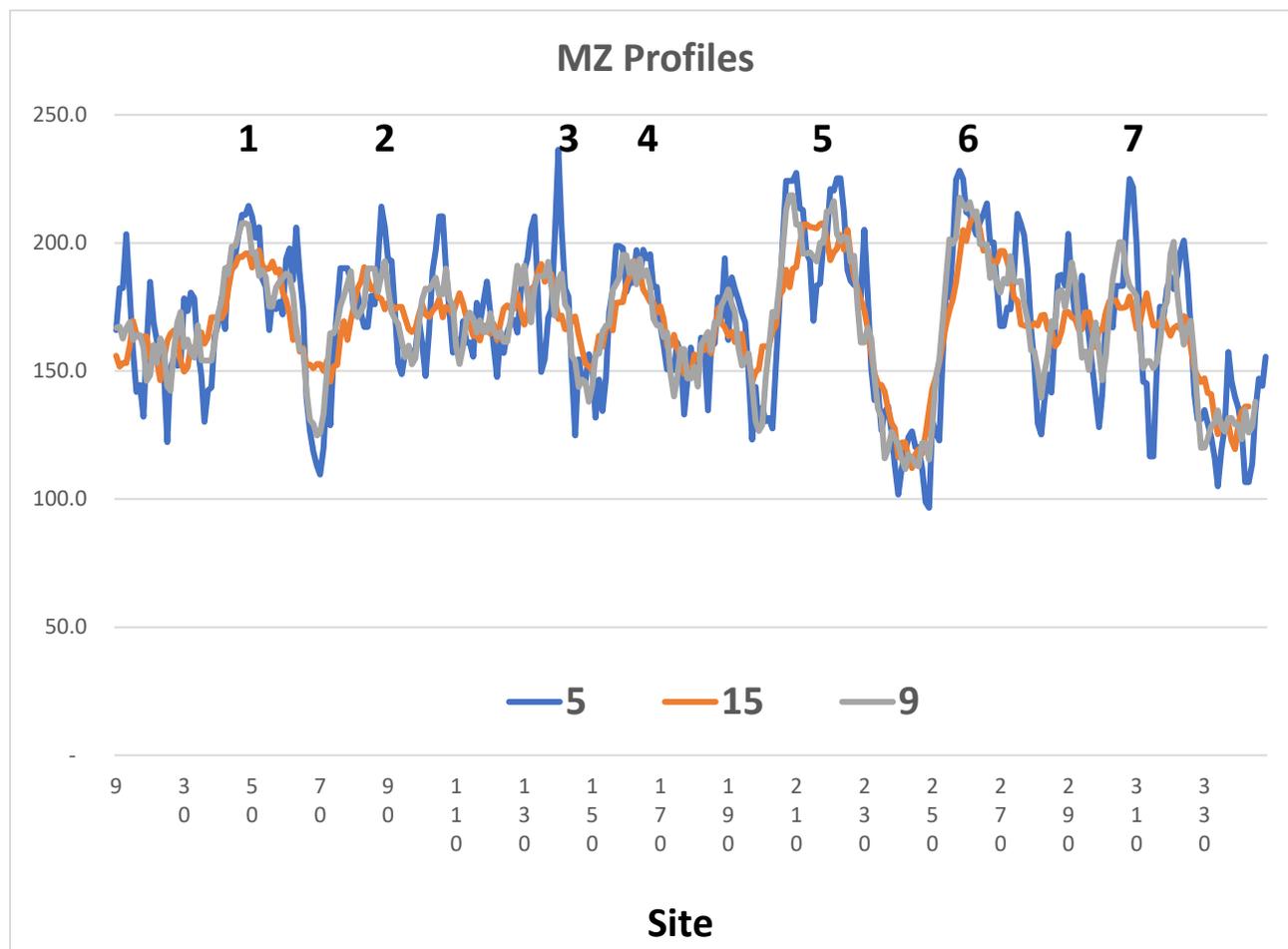

Fig. 3.  The transmembrane regions that correspond to those listed by Uniprot, and here in Table 1, are numbered.  They are associated with hydrophobic peaks in the Ψ(aa,W) profiles for W = 5, 9, and 15.

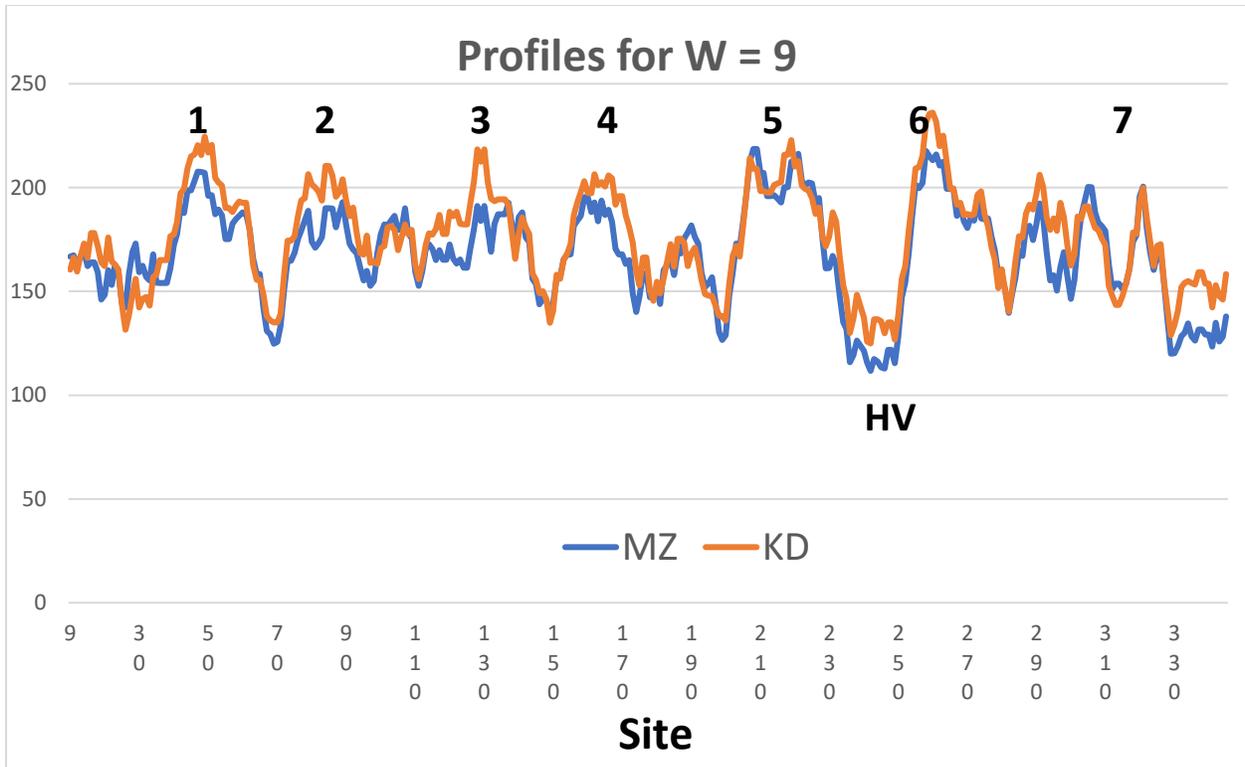

Fig. 4. Here the MZ and KD Ψ(aa,W) profiles for W = 9 show small and systematic differences.

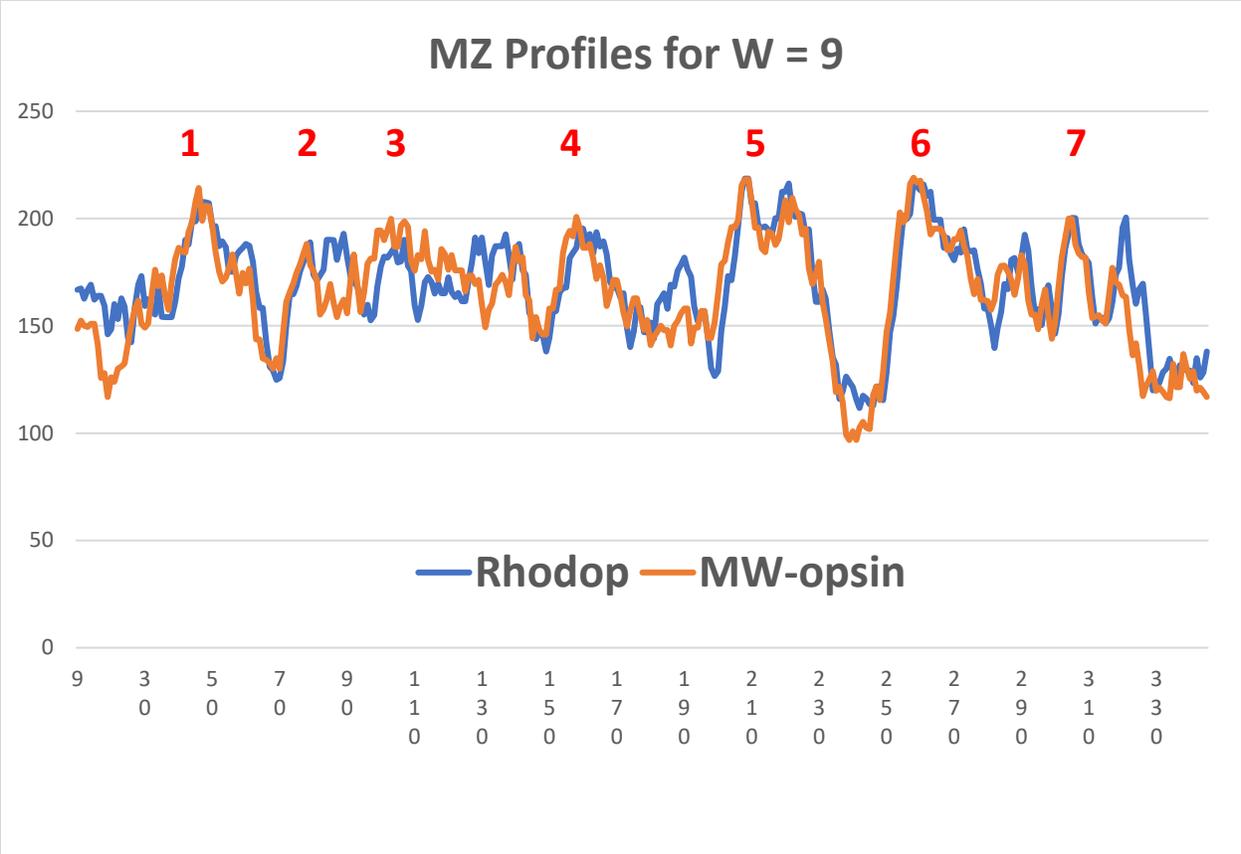

Fig. 5.

|   | W = 9 | Uniprot P08100 | Helix | |
|---|---|---|---|---|
| 1 | 49 | 49 | 35-64 | |
| 2 | 84 | 85 | 74-88 | 90-100 |
| 3 | 134 | 122 | 106-140 | |
| 4 | 165 | 163 | 143-168 | 170-173 |
| 5 | 216 | 213 | 200 - | |
| 6 | 259 | 263 | 306 | |
| 7 | 305 | 297 | 311-321 | |

Table 1. The centers of the transmembrane regions identified in Fig. 4 are compared to those listed on Uniprot, as well as the helical regions, also listed on Uniprot from structural studies.